\documentclass[prl,twocolumn,showpacs]{revtex4}
\usepackage[dvips]{graphicx}
\usepackage{latexsym}
\usepackage{bm}
\begin{document}
\newcommand{\fig}[2]{\includegraphics[width=#1]{#2}}

\title{Nodal $d+id$ pairing and topological phases on the triangular lattice: \\
unconventional superconducting state of Na$_x$CoO$_2\cdot y$H$_2$O}
\author{Sen Zhou$^{1,2}$ and Ziqiang Wang$^1$}
\affiliation{$^1$Department of Physics, Boston College, Chestnut
Hill, MA 02467} \affiliation{$^2$National High Magnetic Field
Laboratory, Florida State University, Tallahassee, FL 32310}

\date{\today}

\begin{abstract}

We show that finite angular momentum pairing chiral superconductors
on the triangular lattice have point zeroes in the complex gap
function. A topological quantum phase transition takes place through
a nodal superconducting state at a specific carrier density $x_c$
where the normal state Fermi surface crosses the isolated zeros. For
spin singlet pairing, we show that the second nearest neighbor
$d+id$-wave pairing can be the dominant pairing channel. The gapless
critical state at $x_c\simeq0.25$ has six Dirac points and is
topologically nontrivial with a $T^3$ spin relaxation rate below
$T_c$. This picture provides a possible explanation for the
unconventional superconducting state of Na$_x$CoO$_2\cdot y$H$_2$O.
Analyzing a pairing model with strong correlation using the
Gutzwiller projection and symmetry arguments, we study these
topological phases and phase transitions
as a function of Na doping. \typeout{polish abstract}
\end{abstract}


\maketitle

The sodium cobaltate Na$_x$CoO$_2$, a layered triangular lattice
electron system, has received widespread interest since the recent
discovery of the 5K superconducting (SC) phase in water intercalated
Na$_x$CoO$_2\cdot y$H$_2$O near $x=0.3$ \cite{takada03}. While a
very rich phase diagram is emerging for a broad range of Na
concentrations \cite{foo04}, the nature of the superconducting phase
has remained poorly understood. The main evidence that the SC phase
is unconventional comes from the absence of the coherence peak in
the in-plane NMR spin-lattice relaxation rate at $T_c$ and its power
law temperature dependence ($T^3$) below $T_c$
\cite{fujimoto04,ishida03,zheng-jpcm06}. Despite this evidence for
an anisotropic SC gap function with line nodes, consistent with the
specific heat \cite{hdyang05} and $\mu$SR \cite{kanigel04}
measurements, there has been considerable debate over the pairing
symmetry. Measurements of angle-averaged Knight shift in powdered
samples have produced inconsistent results supporting either
spin-triplet \cite{kato-jpcm06} or spin-singlet pairing
\cite{kobayashi05k}. The most recent measurements on high quality
single crystals \cite{zheng06k} show that the spin contributions to
the Knight shift decreases below $T_c$ along both the $a$ and
$c$-axis, which strongly supports that the Cooper pairs are formed
in the spin-singlet state. The Knight shift and the relaxation rate
above $T_c$ show antiferromagnetic (AF) correlations
\cite{zheng06k,zheng-jpcm06}.

On the theoretical side, there has been a growing number of
proposals for unconventional superconductivity in the cobaltates.
For spin-singlet pairing in a non-s-wave channel, the six-fold
symmetry of the triangular lattice requires the low angular momentum
paired state to have the chiral $d_{x^2-y^2}\pm id_{xy}$ symmetry,
raising the exciting possibility of a time-reversal symmetry
breaking superconductor. Earlier studies that drew analogy to the
high-T$_c$ cuprates indeed found unanimously spin-singlet, $d+id$
pairing via the nearest neighbor AF superexchange in the triangular
lattice $t$-$J$ model at low doping
\cite{baskaran03,kumar03,ogata03,wangleelee04}. However, it is
conventional wisdom that a $d+id$ paired state, as well as other
chiral SC states with complex order parameters, has a full gap and
is thus inconsistent with NMR experiments
\cite{fujimoto04,ishida03,zheng-jpcm06}.

In this paper, we show that extended chiral paired states beyond the
nearest neighbor (NN) have generic point zeroes in the complex gap
function {\em inside} the first Brillouin zone. Consider
$(\ell,n)$-wave pairing with angular momentum $\ell$ for the pairs
on the $n$-th NN bond, the order parameter can be written down in
real space,
\begin{equation}
\Delta_{ij}=\Delta_{\ell n} e^{i\ell\theta_{ij}},
\end{equation}
where $\ell$ labels the angular momentum of the pair and
$\theta_{ij}$ is the angle of ${\vec r_{ij}}={\vec r_i}-{\vec r_j}$
between sites $i$ and $j$. The Fourier transform of $\Delta_{ij}$ is
given by $\Delta_{\ell n}(k)=2[\beta_{\ell n}^\prime(k)+i\beta_{\ell
n}^{\prime\prime}(k)]$, where the real ($\beta^\prime$) and the
imaginary ($\beta^{\prime\prime}$) parts of the gap function are
given in Table I in terms of the triangular lattice harmonics for
the chiral $p$-wave ($\ell=1$) and $d$-wave ($\ell=2$) pairing.
Nodes in the complex gap function arise where the real and imaginary
parts of $\Delta_{\ell n}(k)$ vanish simultaneously, \textit{i.e.}
at the crossing points of the lines of zeroes of $\beta^\prime(k)$
and $\beta^{\prime\prime}(k)$. Fig. 1 shows the locations of the gap
nodes in the first zone for chiral $p$ and $d$-wave pairing. For
first NN pairing, the nodes are pinned to the zone center and zone
boundary and thus a generic chiral $(\ell, 1)$-wave superconducting
state has a full gap. Remarkably, for the case of $n>1$, new nodes
in the gap function appear {\em inside} the zone and can support
nodal $(\ell,n>1)$-wave superconducting states.

Hereafter we focus on the spin-singlet $d+id$-wave case,
\textit{i.e.} chiral $(2,n)$-paired states and argue that the
existence of the nodes is relevant for understanding the
superconducting state in hydrated cobaltates. For the second NN
pairing with $n=2$, there are six nodes inside the first zone marked
by the solid circles in Fig.~1b. They are located at $\pm{\bf
k}^*_\alpha$, with $\alpha=1,2,3$, and ${\bf
k}^*_1=2\pi[2/(3\sqrt{3}),0]$, ${\bf
k}^*_2=2\pi[1/(3\sqrt{3}),1/3]$, and ${\bf
k}^*_3=2\pi[1/(3\sqrt{3}),-1/3]$, coinciding with the six corners of
the hexagonal $\sqrt{3}\times\sqrt{3}$ reduced zone boundary.

\begin{table}[htb]
\begin{ruledtabular}
\text{ Chiral {\textit{p}}-wave pairing ($\ell=1$):}
\begin{tabular}{ccc}
$n$&$\beta_{1,n}^\prime(k)$&$\beta_{1,n}^{\prime\prime}(k)$ \\
\hline 1 & $\sqrt{3}\sin{{\sqrt{3}\over 2}k_x}\cos{{1\over 2}k_y}$
& $\sin{k_y}+\cos{{\sqrt{3}\over 2}k_x}\sin{{1\over 2}k_y}$\\
2 & $-\sqrt{3}\sin{{3\over 2}k_y}\cos{{\sqrt{3}\over 2}k_x}$
& $\sin{\sqrt{3}k_x}+\cos{{3\over 2}k_y}\sin{{\sqrt{3}\over 2}k_x}$\\
3 &$\sqrt{3}\sin{\sqrt{3}k_x}\cos{k_y}$
& $\sin{2k_y}+\cos{\sqrt{3}k_x}\sin{k_y}$\\
\end{tabular}
\vspace{0.5cm} \text{ Chiral {\textit{d}}-wave pairing ($\ell=2$):}
\begin{tabular}{ccc}
$n$&$\beta_{2,n}^\prime(k)$&$\beta_{2,n}^{\prime\prime}(k)$ \\
\hline 1 & $\cos{k_y}-\cos{{\sqrt{3}\over 2}k_x}\cos{{1\over 2}k_y}$
& $\sqrt{3}\sin{{\sqrt{3}\over 2}k_x}\sin{{1\over 2}k_y}$\\
2 & $\cos{\sqrt{3}k_x}-\cos{{3\over 2}k_y}\cos{{\sqrt{3}\over
2}k_x}$
& $-\sqrt{3}\sin{{3\over 2}k_y}\sin{{\sqrt{3}\over 2}k_x}$\\
3 &$\cos{2k_y}-\cos{\sqrt{3}k_x}\cos{k_y}$
& $\sqrt{3}\sin{\sqrt{3}k_x}\sin{k_y}$\\
\end{tabular}
\end{ruledtabular}
\caption{\label{gapk} Gap function $\Delta_{\ell
n}(k)=2\left[\beta'_{\ell n}(k)+i\beta''_{\ell n}(k)\right]$.}
\vskip-0.3cm
\end{table}

The first indication that this is special for the triangular lattice
cobaltates comes from the fact that a single hexagonal Fermi surface
(FS) at the SC concentration $x=1/3$ matches the
$\sqrt{3}\times\sqrt{3}$ reduced zone boundary such that these nodes
would lie directly on the FS. Recent angle-resolved photoemission
(ARPES) experiments in the both the unhydrated \cite{hbyang05,hasan}
and the hydrated \cite{shimojima} compounds near $x\sim 0.3$ indeed
observe a single rounded hexagonal FS that coincides well with the
hexagonal $\sqrt{3}\times\sqrt{3}$ reduced zone boundary. This is
highly unexpected since the Co$^{4+}$ has $5$ $d$-electrons
occupying the three $t_{2g}$ orbitals. It turns out that strong
correlation effects beyond the band theory renormalize the crystal
field splitting and bandwidths, leading to a single quasiparticle
band crossing the Fermi level as observed by ARPES experiments
\cite{zhou05}. In Fig.~1b, we show the FS calculated in
Ref.~\cite{zhou05} at $x_c=0.25$, which passes through the six gap
nodes for the second NN $d+id$-wave pairing. The value of $x_c$ is
smaller than 1/3 due to the rounding of the hexagonal FS in both the
ARPES data \cite{hbyang05} and the theoretical calculations
\cite{zhou05}. Interestingly, third NN $d+id$-wave pairing
introduces six nodes (Fig.~1b) at the corners of the hexagonal
$2\times2$ reduced zone, which intersects the FS at $x=1/2$ where
the cobaltate is in an insulating phase with charge and spin order
\cite{foo04,zhou07}.

\begin{figure}
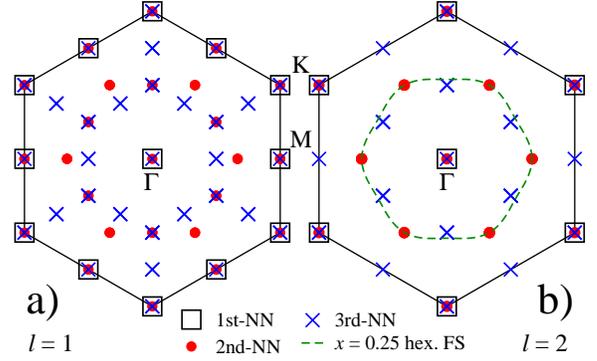

\begin{center}
\fig{3.0in}{fig1.eps} \caption{Nodes of chiral $(\ell,n)$-wave
pairing where the gap function $\Delta_{\ell,n}(k)$ vanishes. (a)
$p$-wave and (b) $d$-wave with normal state FS at $x=0.25$ (dashed
line).} \label{fig1}
\end{center}
\vskip-0.5cm
\end{figure}

The second, direct evidence for nodal $d+id$-wave pairing at $x=x_c$
comes from the most recent NMR experiments. In high quality single
crystals, Zheng \textit{et. al.} discovered that the $T^3$ decay of
the relaxation rate $1/T_1$, \textit{i.e.} the signature of line
nodes in the gap function, is obeyed down to the lowest temperatures
only at $x_c\simeq0.26$, whereas deviations from $T^3$ are observed
on both sides of $x_c$ \cite{zheng-jpcm06}. The existence of the
Dirac nodes in the gap function at the critical doping $x_c$ for
spin-singlet pairing on the triangular lattice strongly favors the
scenario of second NN $d+id$-wave pairing. In the following, we
report a variational study of the $(2,n)$-wave pairing states in an
effective single-band $t-U$ plus pairing model of the cobaltates,
calculate the Knight shift and the spin relaxation rate $1/T_1$, and
describe the topological properties of the superconducting phases.

The Hamiltonian is given on a triangular lattice,
\begin{equation}
H=\sum_{ij,\sigma}t_{ij}c^{\dagger}_{i\sigma} c_{j\sigma}
+U\sum_{i}{\hat n}_{i\uparrow}{\hat n}_{i\downarrow}
-\sum_{i,j}W_{ij}{\hat\Delta}_{ij}^\dagger{\hat\Delta}_{ij},
\label{hrs}
\end{equation}
where $c^{\dagger}_{i\sigma}$ creates a hole of spin $\sigma$ and
$U$ is the Co on-site Coulomb repulsion. To describe the $a_{1g}$
band crossing the Fermi level, we fix the first three NN hopping
$t_{ij}=(t_1,t_2,t_3)=(-202, 35, 29)$ meV \cite{zhou07}. The hole
density $n_i=1-x_i$ where $x_i$ is the electron doping
concentration. The last term in Eq.~(\ref{hrs}) is a
phenomenological singlet pairing interaction $W_{ij}>0$ between
$n$-th NN sites $i$ and $j$ with ${\hat\Delta}_{ij}=
c_{i\uparrow}c_{j\downarrow}-c_{i\downarrow}c_{j\uparrow}$. The
effects of strong correlation are accounted for in the Gutzwiller
approximation \cite{ga}, leading to a renormalized band dispersion,
\begin{eqnarray}
\xi_{k}&=& 2g_tt_1(\cos k_y+2\cos\sqrt{3}k_x/{2}\cos{k_y}/{2})
\nonumber \\
&+& 2g_tt_2(\cos\sqrt{3}k_x+2\cos\sqrt{3}k_x/{2}\cos {3k_y}/{2})
\nonumber \\
&+& 2g_tt_3(\cos2k_y+2\cos\sqrt{3}k_x\cos k_y)-\mu, \label{ek}
\end{eqnarray}
where $g_t=2x/(1+x)$ is the Gutzwiller renormalization factor in the
large-U limit \cite{zhou07,ga}. The BCS state for $(2,n)$-wave
pairing can be written equivalently as \cite{readgreen00}
\begin{equation}
|\Psi_{n}\rangle=\prod_k|u_{nk}|^{1/2}\exp \left({1\over2}
g_{nk}c_{k\uparrow}^\dagger c_{-k\downarrow}^\dagger
\right)|0\rangle, \label{wavefunction}
\end{equation}
where $g_{nk}=v_{nk}/u_{nk}=-(E_{nk}-\xi_k)/\Delta_{nk}^*$ and the
quasiparticle excitation energy
$E_{nk}=\sqrt{\xi_k^2+|\Delta_{nk}|^2}$. The gap function is
$\Delta_{nk}=2W_n\Delta_n[\beta_{2,n}^\prime(k)+i\beta_{2,n}^{\prime\prime}(k)]$,
with $\beta$ given in Table I. We first study the variational ground
state obtained by minimizing the energy with respect to the only
variational parameter $\Delta_n$ for a given $W_n$. In Fig.~2a, we
compare $\Delta_n$ for the 1st, 2nd, and 3rd NN pairing with
identical pairing strength $W_{1,2,3}=50$meV at several dopings. The
2nd NN $d+id$ pairing is the strongest with the largest pairing
order parameter. Including a NN Coulomb repulsion $V$ \cite{mlee}
would further suppress the 1st NN pairing attraction $W_1$, making
the 2nd NN $d+id$-wave the dominant spin-singlet pairing channel
that drives the SC transition. We emphasize that while the
subdominant channels can emerge below the transition, the admixture
of a small component with $n\ne2$ would only result in small shifts
of the six-fold symmetric point nodes in the gap function and not
change the results discussed here.

\begin{figure}
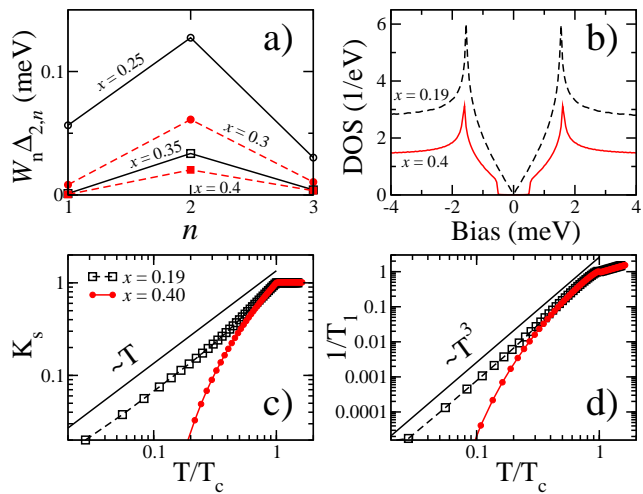

\begin{center}
\fig{3.3in}{fig2.eps} \caption{(a) Comparison of $d+id$ pairing
order parameters at several dopings for $W_{1,2,3}=50$ meV. (b-d)
Properties of 2nd NN $d+id$-wave pairing at $x=0.19=x_c$ for
$W_2=50$meV and $x=0.4>x_c$ for $W_2=73$meV: Tunneling DOS (b),
temperature dependence of Knight shift $K_s$ (c) and NMR relaxation
rate $1/T_1$ (d) normalized by their values at $T_c$.} \label{fig3}
\end{center}
\vskip-0.5cm
\end{figure}

Next we turn to the properties of the second NN $d+id$ state. For
our band parameters, the FS is almost circular and $x_c\simeq 0.19$.
At $x=x_c$, the normal state FS crosses the six nodes at $\pm{\bf
k}_\alpha^*$ around which the quasiparticle dispersion $E_k$
(hereafter we drop the index $n$) has a conical spectrum. For
example, expanding around $\pm{\bf k}^*_1$,
$E_k\simeq\sqrt{(A^2+B^2)[k_x\mp 4\pi/(3\sqrt{3})]^2+B^2k_y^2}$,
with $A=3(t_1-2t_3)$ and $B=9W\Delta/2$. Thus, the six nodes are
described by three pairs of Dirac fermion doublets with anisotropic
velocities. They govern the properties of the low energy
excitations. We calculate the tunneling density of states (DOS)
$N(E)$ and the temperature dependence of the $ab$-plane Knight shift
($K_s$) and the spin relaxation rate ($1/T_1$) according to
\begin{equation}
\left\{K_s,{1\over T_1}\right\}\propto-\int_{-\infty}^{\infty}
\left\{1, TN(E)\right\}N(E){\partial f(E)\over \partial E} dE.
\label{nmr}
\end{equation}
The results are plotted in Figs.~2b-d. The calculated $N(E)$ shows a
linearly vanishing, V-shaped DOS at $x=x_c$ generic of nodal
$d$-wave pairing. As a consequence, the Knight $K_s$ and $1/T_1$
follow the $T$ and $T^3$ behaviors below $T_c$ in agreement with the
NMR experiments on single crystals at $x\simeq0.26$
\cite{zheng06k,zheng-jpcm06}. For $x\ne x_c$, the normal state FS
does not overlap the nodes in the gap function. Fig.~2b shows that
at $x=0.4$ the DOS is fully gapped at low energies, but reverts back
to the V-shape above the energy gap. This leads to a rapid crossover
of $K_s$ and $1/T_1$ from the power law behaviors just below $T_c$
to exponential decays at low temperatures (Figs.~2c and 2d).
In this case, we expect that disorder-induced filling of the DOS gap
to produce crossovers to constant $K_s$ and linearly vanishing
$1/T_1$ at low temperatures \cite{fujimoto04}.

We now discuss the topological properties of the chiral SC states
and the topological quantum phase transition as the carrier density
$x$ moves across $x_c$. The topological order in chiral
fermion-paired states has been a subject of growing interests in
connection to unconventional superconductors/superfluids and
fractional quantum Hall states
\cite{volovik97,senthil99,readgreen00}. The key point is that the
complex order parameter $\Delta_k$ and the dispersion $\xi_k$ form a
unit pseudospin vector introduced by Anderson \cite{anderson58}:
${\bf m}(k)=({\rm Re}\Delta_k,-{\rm Im}\Delta_k,\xi_k)/E_k$. Since
$|{\bf m}|^2=1$, ${\bf m}$ lives on a 2-sphere $S^2$. The SC state
can be viewed as a BCS mapping from the ${\bf k}$-space, which is a
torus $T^2$ (compactified from an infinite plane) to the pseudospin
$S^2$. This becomes evident when we write ${\bf m}$ in term of the
pairing function $g_k=v_k/u_k$ in Eq.~(\ref{wavefunction}),
\begin{equation}
m_1(k)+im_2(k)=-{2g_k\over 1+\vert g_k\vert^2}, \quad
m_3(k)={1-\vert g_k\vert^2\over 1+ \vert g_k\vert^2}. \label{w}
\end{equation}
Such maps are classified by homotopy classes with topological
winding numbers $Q$. The lattice version of $Q$ is,
\begin{equation}
Q={1\over8\pi}\sum_{\rm \triangle} {\bf m}(k_1)\cdot[{\bf m}
(k_2)\times{\bf m}(k_3)],\label{topo}
\end{equation}
where the sum is over all elemental triangular plaquettes with
corners labeled as $1,2,3$. Topologically distinct phases are
categorized by different winding numbers $Q$ that count the number
of times the pseudospin configuration ${\bf m}$ wraps around the
sphere. We find that $Q\ne0$ if $g_k$ has a {\em singular} vortex
structure near the nodes of the complex gap function, a condition
that depends on the location of the FS. In Fig.~3, we show the
calculated $Q$ for $x>x_c$, $x<x_c$, and at the critical state
$x=x_c$. All of them are topologically nontrivial. For $x>x_c$, the
normal state FS encloses only the gap node at the zone center where
$g_k\propto1/(k_x-ik_y)^2$, exhibiting a singular double vortex.
This charge-two magnetic monopole contributes two flux quanta to the
surface integral and gives $Q=\mp2$ (Fig.~3) for the fully gapped
topological phase. The sign corresponds to the relative sign of the
real and imaginary parts of the order parameter. Note that near the
nodes {\em outside} the FS, $g_k$ behaves as zeroes, e.g.
$g_k\propto k_x\mp4\pi/(3\sqrt{3})+ik_y$ near $\pm{\bf k}^*_1$, and
thus do not contribute to $Q$. As the doping is reduced to $x< x_c$,
the normal state FS encloses, in addition to the node at $\Gamma$,
the six nodes at $\pm{\bf k}_\alpha^*$ where $g_k$ exhibits singular
vortex structure; e.g. $g_k\propto 1/[k_x\mp4\pi/(3\sqrt{3})+ik_y]$
near $\pm{\bf k}^*_1$. Each pair of the nodes can be viewed as a
singular double vortex. The contribution to the magnetic flux
through the sphere is thus increased by six flux quanta leading to
$Q=\pm4$ as shown in Fig.~3.

In the critical state at $x=x_c$, the FS passes through the six gap
nodes. The pseudospin vector ${\bf m}$ is not well defined exactly
at these isolated diabolic points, which are lattice-regulated in
the summation in Eq.~(\ref{topo}). Remarkably, Fig.~3 shows that the
winding number $Q=\pm1$ and the critical state is a gapless
topologically nontrivial state. This suggests that each pair of the
gap nodes in the critical state contributes a single vortex in
$k$-space, reminiscent of the situation in the $p+ip$ state
\cite{volovik88,readgreen00}. Indeed, evaluating $Q$ in the
continuum limit along the line integrals of the six circles cutting
out the nodes at ${\bf k}_\alpha^*$, one finds that each of the
latter contributes half a magnetic flux leading to the total
$Q=\pm1$. We note that the existence of such a {\em gapless}
topological phase in two dimensions is new and consistent with the
view point that gapless fermions emerge as a result of a form of
quantum order \cite{wenzee} in the critical state. Physically, the
topologically invariant winding number $Q$ corresponds to the
quantization of the spin Hall conductance $\sigma_{xy}^s$
\cite{senthil99,readgreen00}. In unit of the spin conductance
quantum $(\hbar/2)^2/2\pi\hbar$, the spin Hall conductance is
$\sigma_{xy}^s=Q$. Thus, we predict that as the electron doping $x$
evolves across $x_c$, the superconducting state of the cobaltate
exhibits quantum spin Hall transitions from $\mp2$ to $\pm4$ with
quantized critical spin Hall conductance $\pm1$. We expect novel
properties associated with the edge states.

\begin{figure}
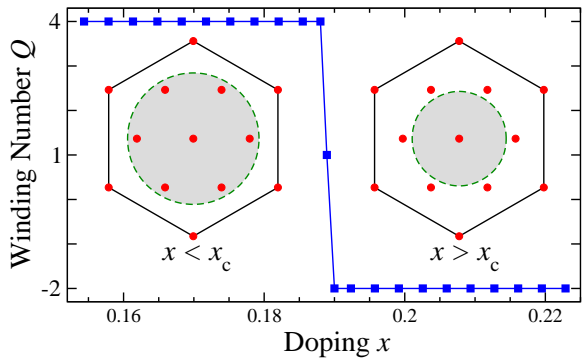

\begin{center}
\fig{3.0in}{fig3.eps} \caption{Topological winder number as a
function of doping in the second NN $d+id$-wave superconducting
state. Insets show the locations of the nodes in the complex gap
function and the normal state FS.} \label{fig4}
\end{center}
\vskip-0.5cm
\end{figure}

In summary, we have proposed a class of nodal chiral
superconductors. The spin-singlet, second NN $d+id$-wave pairing
turns out to be consistent with the NMR experiments on high quality
single crystals of hydrated cobaltate superconductors, and has some
remarkable topological properties related to the quantum spin Hall
transitions. Although the microscopic origin for the second NN
$d+id$ pairing is beyond our scope here, it is conceivable that (i)
a strong NN Coulomb repulsion \cite{mlee}, (ii) a larger 2nd NN
superexchange, and (iii) the frustration of the NN antiferromagnetic
correlation on the triangular lattice and the proximity to
inhomogeneous charge/spin ordered state \cite{mlee,zhou07} will
favor a dominant extended pairing interaction beyond the first NN.
It is noted that the chiral pairing state breaks time-reversal
symmetry which should in principle be detectable by $\mu$SR or
optical Kerr experiments with sufficient resolution. The orbital
current near a unitary impurity produces a static magnetic field. To
estimate the size of the field, we performed self-consistent
calculations for the circulating current near a unitary impurity.
The maximum current around the impurity site is $\sim100$nA for a
pairing strength $W_2=50$meV corresponding to a $T_c\sim5K$. This
gives an estimate of the magnetic field at the impurity site
$\sim1$G, which is close to the upper bound set by the earlier
$\mu$SR experiments \cite{higemoto04,uemura04}. Future experiments
are very desirable to determine whether time-reversal symmetry is
broken in the superconducting cobaltates.

We are grateful to Y. Yu for many useful discussions. This work was
supported by DOE grant DE-FG02-99ER45747 and NSF grant DMR-0704545.

\end{document}